\documentstyle[epsfig,prl,aps,floats]{revtex}	

\begin{document}
\draft	

\title{Whispering Vortices}
\author{A.~Wallraff 
\and A.~V.~Ustinov
\and V.~V.~Kurin\cite{address}
\and I.~A.~Shereshevsky\cite{address}
\and N.~K.~Vdovicheva\cite{address}}

\address{Physikalisches Institut III, Universit{\"a}t Erlangen-N{\"u}rnberg, 
D-91058 Erlangen, Germany} 

\date{\today}

\wideabs{ 

\maketitle

\begin{abstract}
Experiments indicating the excitation of whispering gallery type 
electromagnetic modes by a vortex moving in an annular Josephson 
junction are reported.  At relativistic velocities the Josephson 
vortex interacts with the modes of the superconducting stripline 
resonator giving rise to novel resonances on the current-voltage 
characteristic of the junction.  The experimental data are in good 
agreement with analysis and numerical calculations based on the 
two-dimensional sine--Gordon model.
\end{abstract}

  
} 
  

Whispering gallery modes are universal linear excitations of circular
and annular resonators.  They have first been observed in form of a
sound wave traveling along the outer wall of a walkway in the circular
dome of St.~Paul's Cathedral in London and were investigated by Lord
Rayleigh \cite{Rayleigh14} and others \cite{Walker78}.  In the 2 meter
wide walkway, which forms a circular gallery of 38 meter diameter
about 40 meters above the ground of the cathedral, the whispering of a
person can be transmitted along the wall to another person listening
to the sound on the other side of the dome.  The investigations by
Rayleigh led to the conclusion that the whisper of a person does
excite acoustic eigenmodes of the circular dome which can be described
using high order Bessel functions.  This acoustic phenomenon lends its
name ``whispering gallery mode'' to a number of similar, mostly
electromagnetic excitations in circular resonators.  Whispering
gallery modes are of strong interest in micro-resonators used for
ultra small lasers \cite{McCall91}.  Most recently, circular
resonators with small deformations, in which chaotic whispering
gallery modes were observed, attracted a lot of attention
\cite{Gmachl98}.  Here we describe the experimental observation of
electromagnetic whispering gallery modes excited by a vortex moving in
an annular Josephson junction of diameter less than $100 \, \rm{\mu
m}$.

A long Josephson junction is an intriguing nonlinear wave propagation 
medium for the experimental study of the interaction between linear 
waves and solitons \cite{Ustinov98}.  In this letter we report the 
excitation of whispering gallery type electromagnetic modes by a 
topological soliton (Josephson vortex) moving at relativistic 
velocities in a wide annular Josephson junction.  We make use of the 
same Josephson vortex for both exciting and detecting the whispering 
gallery mode.  These modes are manifested by their resonant 
interaction with the moving vortex, which results in a novel fine 
structure on the current-voltage characteristic of the junction.  Our 
experiments are consistent with the recently published theory 
\cite{Kurin98} based on the two-dimensional sine-Gordon model.  We 
present numerical calculations based on this model, which show good 
agreement with experiments.
\begin{figure}[tb]
 	\centering
    \epsfig{file=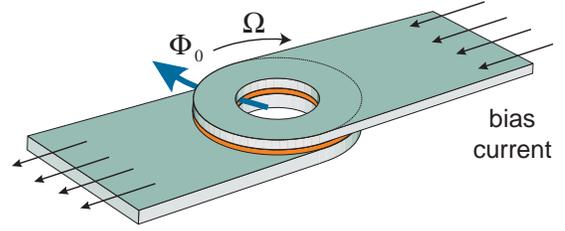, width=3.0 in}
	\caption{
	Geometry of the annular Josephson junction.  The direction of the 
	angular velocity $\Omega$ of the vortex ($\Phi_{0}$) under the action 
	of the bias current is indicated.
	}
	\label{fig:ring_geom_top}
\end{figure}

Electromagnetic waves in an annular Josephson junction are described 
by the perturbed sine-Gordon equation (PSGE) for the superconducting 
phase difference $\phi$ between the top and bottom superconducting 
electrodes of the junction \cite{Ustinov98}.  The Josephson vortex, 
often also called fluxon, corresponds to a twist over $2 \pi$ in 
$\phi$.  It carries a magnetic flux equal to the magnetic flux 
quantum $\Phi_{0}=h/2e = 2.07\,10^{-15}\,\rm{Vs}$.  Physically, this 
flux is induced by a vortex of the screening current flowing across 
the junction barrier.  Linear excitations in this system are Josephson 
plasma waves that account for small amplitude oscillations in $\phi$.
The maximum phase velocity of electromagnetic waves in such a junction 
is the Swihart velocity given by $c_{0} = \lambda_{J} \, 
\omega_{p}$, where $\lambda_{J}$ is the Josephson length and 
$\omega_{p}$ the plasma frequency \cite{Ustinov98}.  In zero external 
magnetic field the PSGE for an annular Josephson junction of width $w 
< \lambda_{J}$ can be written as
\begin{equation}
      \left( \nabla^{2}
	- \frac{\partial^2}{\partial t^2}\right) \phi
	- \sin{\phi} = \gamma + \alpha \frac{\partial \phi}{\partial t}
	- \beta \nabla^2 \frac{\partial \phi}{\partial t},
\label{eq:2DsineGordon}
\end{equation}
where space and time are normalized by $\lambda_{J}$ and
$\omega_{P}^{-1}$, respectively.  In Eq.~(\ref{eq:2DsineGordon})
$\nabla^{2} - \partial^2/\partial t^2$ is the D'Alembert wave
operator, $\sin \phi$ is the nonlinear term due to the phase-dependent
Josephson current and $\gamma$ is the normalized bias current.  The
damping terms $\alpha \, \partial \phi/\partial t$ and $\beta \,
\nabla^2 \partial \phi/\partial t$ are inversely proportional to the
quasiparticle resistance across the junction barrier and to the
quasiparticle impedance of the electrodes, respectively.  For the
junctions of width $w<\lambda_{J}$ considered in this paper, a
homogeneously distributed bias current $\gamma$ as in
Eq.~(\ref{eq:2DsineGordon}) is justified.  In contrast, for junctions
with $w > \lambda_{J}$ the bias current may contribute to the
boundary conditions of Eq.~(\ref{eq:2DsineGordon}) \cite{Martucciello96b}.

A vortex steadily moving at a velocity $u$ driven by the Lorentz 
force due to the bias current $\gamma$ generates a voltage $V \propto 
u$ across the Josephson junction.  This voltage can be monitored in 
experiment.  The radiation associated with the time-dependent fields 
described by Eq.~(\ref{eq:2DsineGordon}), i.e.~the magnetic field $H 
\propto | \nabla \phi |$ and the electric field $E \propto 
\partial \phi / \partial t$, can be measured either directly (for certain 
junction geometries) or through its interaction with the moving 
vortex.

In contrast to most of the previous experiments focusing on
quasi-one-dimensional annular Josephson junctions, we investigate
comparatively wide, effectively two-dimensional junctions.  We have
fabricated a set of 5 annular Josephson junctions (A $\ldots$ E) with
the ratio $\delta = r_{i}/r_{e}$ between the inner radius $r_{i}$ and
the fixed outer radius $r_{e} = 50 \, \rm{\mu m}$ being varied between
$\delta = 0.94$ and $\delta = 0.60$ (see Table~\ref{tab:parameters}). 
The junctions are made at Hypres Inc.\cite{Hypres} using
Nb-Al/AlO$_{x}$-Nb trilayer technology and employ the standard biasing
geometry \cite{Davidson85} as shown in Fig.~\ref{fig:ring_geom_top}. 
Due to the fabrication technology, the junction area is surrounded by
a small passive region about $2 \, \rm{\mu m}$ wide, which is omitted
from Fig.~\ref{fig:ring_geom_top} for clarity.  In the passive region
the top and bottom electrodes are separated by a $200 \, \rm{nm}$
thick SiO$_{2 }$ layer, which act as a small stripline in parallel to,
but with electrical parameters different from the junction itself.

\begin{table}[tb]
	\centering
	\caption{Geometrical parameters of the annular Josephson junctions 
	used in experiment. The outer radius of every junction is 
	$r_{e}=50\,\rm{\mu m}$. $w = r_{e}-r_{i}$ is the junction width.}
	\begin{tabular}{lccccc}
		junction & A & B & C & D & E \\
		\hline
		$r_{i}$ [$\mu$m]& 47 & 45 & 42 & 35 & 30 \\
		$w$ [$\mu$m]& 3 & 5 & 8 & 15 & 20 \\
		$\delta = r_{i} / r_{e}$ & 0.94 & 0.90 & 0.84 & 0.70 & 0.60 \\
		$\xi = c_{0} / \bar{c}_{0}$ & 0.70 & 0.78 & 0.85 & 0.91 & 0.94 \\		
	\end{tabular}
	\label{tab:parameters}
\end{table}

All junctions show a homogeneous bias current distribution, inferred
from the large value of the vortex-free critical current at zero
field, which is close to the theoretical limit.  Their critical
current density is $j_{c} \approx 160 \, \rm{A cm^{-2}}$ and the
London penetration depth is $\lambda_{L}\approx 90 \, \rm{nm}$ at
$4.2\,K$. The thicknesses of the top and the bottom
superconducting electrode are both well in excess of
$\lambda_{L}$.  Accordingly, the characteristic parameters are
estimated as $\lambda_{J} \approx 30 \, \rm{\mu m}$ and $\nu_{p}
\equiv \omega_{p}/2\pi \approx 50 \, \rm{GHz}$.  All presented
measurements were done at $T = 4.2 \,\rm{K}$ using a well shielded low
noise measurement setup.

We could realize single and multiple vortex states repeatedly and
reproducibly in any of the junctions.  Vortices were trapped by
applying a small bias current during cooling down from the normal to
the superconducting state.  Single-vortex states are identified as the
lowest quantized voltage step observed on the current-voltage
characteristics.  Also, a characteristic change of the critical
current modulation with magnetic field accompanied by a suppression of
the critical current by a factor of more than 100 at zero field, as
reported earlier \cite{Vernik97}, was observed when a vortex was
trapped in the junction.

\begin{figure}[tb]
 	\centering
    \epsfig{file=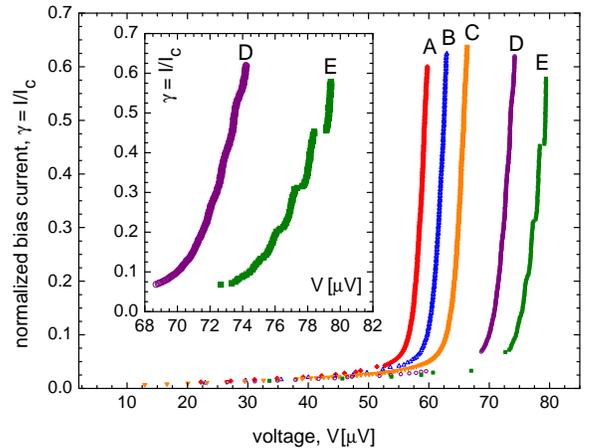, width=3.0 in}
    \caption{Experimental normalized current-voltage characteristics of
    single-vortex states in junctions A to E. An enlargement of the high
    voltage region of the resonances in junctions D and E is shown in the
    inset.
     }
	\label{fig:fluxon_norm_idle}
\end{figure}

In Figure~\ref{fig:fluxon_norm_idle} the single-vortex characteristics
of the junctions A to E are shown.  The current scale is normalized by
the flux-free zero-field critical current of each junction.  The
voltage of each characteristic is multiplied by a factor $\xi =
c_{0}/\bar{c}_{0}$ (see Tab.~\ref{tab:parameters}) calculated using
the approach by Lee et al.~\cite{Lee92}, in order to substract the
effect of a small passive region.  $c_{0}$ ($\bar{c}_{0}$) is the wave
velocity in the junction neglecting (including) the passive region.

A striking novel feature noticed in Fig.~\ref{fig:fluxon_norm_idle} is
the fine structure on the vortex resonance which appears with
increasing the junction width.  The fine structure is most clearly
visible for the widest junction E (see inset of
Fig.~\ref{fig:fluxon_norm_idle}).  We argue that the observed fine
structure can be well understood as due to the interaction of the
moving vortex with the linear whispering gallery modes \cite{Kurin98}. 
The linear modes of the annular Josephson junction resonator are
solutions to the inhomogeneous D'Alembert equation in the polar
coordinates
$(r,\,\theta)$
\begin{equation}
       \left (\frac{1}{r} \frac{\partial}{\partial r}r \frac{\partial}
	   {\partial r} + \frac{1}{r^2}\frac{\partial^2}
	   {\partial \theta^2} - \frac{\partial^2}
	   {\partial t^2} - 1\right) \phi^{(\rm{lin})} = 0,
\label{eq:2DLineareq}
\end{equation}
which is found from Eq.~(\ref{eq:2DsineGordon}) neglecting all
perturbations ($\gamma$, $\alpha \, \partial/\partial t$, $\beta
\nabla^2 \partial \phi/\partial t$) and approximating
the nonlinearity as $\sin \phi \approx \phi$ to take into account the gap
in the plasmon excitation spectrum.  In zero external magnetic field
the boundary conditions
\begin{equation}
	\frac{\partial \phi^{(\rm{lin})}}{\partial r}(r = r_{i},r_{e}) = 0
	\label{eq:2DBoundCond}
\end{equation}
have to be fulfilled.  In terms of the electromagnetic waves in the 
junction, Eq.~(\ref{eq:2DBoundCond}) corresponds to a total internal 
reflection condition.  For large angular wave numbers $k \gg 1$ and wide 
junctions $\delta \ll 1$ a solution to (\ref{eq:2DLineareq}) is 
given by
\begin{equation}
	\phi^{(\rm{lin})}_{k}(r,\theta,t) = A \, J_{k}(\omega_{k} r)\exp(ik\theta) 
	\exp(i\omega_{k} t) \:\:, 
	\label{eq:2DLinearsol}
\end{equation}
where A is an arbitrary amplitude factor, $J_{k}$ is the Bessel 
function of the first kind and $\omega_{k}$ is the angular frequency 
associated with the mode $k$ satisfying the boundary condition 
(\ref{eq:2DBoundCond}) at the external radius.  By calculating  
the dependence of $\omega_{k}$ on $k$, one obtains the dispersion 
relation of whispering gallery modes in the annular junction.

The resonance condition between the angular frequency $\Omega$ of the 
vortex rotation in the ring and the characteristic frequency 
$\omega_{k}$ of a whispering gallery mode can be expressed as
\begin{equation}
	\Omega = \omega_{k}/{k} \:\: .
	\label{eq:resCond}
\end{equation}
In a dispersion diagram Eq.~(\ref{eq:resCond}) determines the crossing 
points between the straight dispersion line of the vortex 
$\omega^{(\rm{v})} = \Omega \, k$ and the dispersion curve $\omega^{(\rm{lin})} 
= \omega_{k}$ of the linear modes.  At low enough damping, a vortex 
moving at the angular frequency $\Omega$ comes into resonance with a 
whispering gallery mode of wavenumber $k$.  If the spacing in $\Omega$ 
between the resonances for different $k$ is large enough, this effect 
can be observed as fine structure resonances on the single-vortex 
current-voltage characteristic of a wide annular Josephson junction.
The increase of the excited wavenumber $k$ with decrease of the vortex 
velocity is a characteristic feature of the interaction between the 
Josephson vortex and the whispering gallery modes of the junction.  
This property is clearly manifested in our numerical calculations 
presented below.
\begin{figure}[tb]
 	\centering
    \epsfig{file=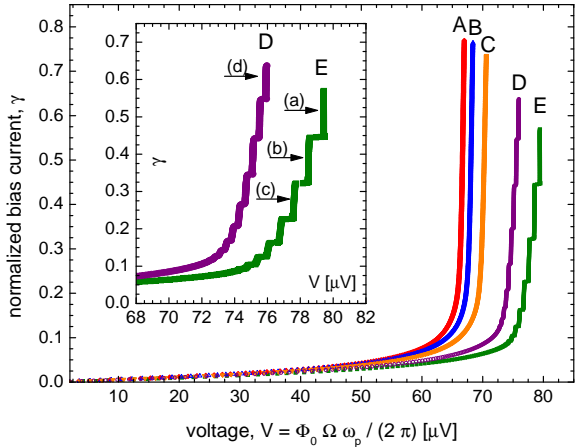, width=3.0 in}
 \caption{Numerically calculated current-voltage characteristics
$V(\gamma)$ for junctions A to E. In the inset the characteristics of
junctions D and E are shown on an enlarged scale.  Arrows indicate the
bias points used to obtain the phase profiles shown in
Fig.~\ref{fig:phaseSim}.}
	\label{fig:IVsim}
\end{figure}

\begin{figure}[tb]
 	\centering
    \epsfig{file=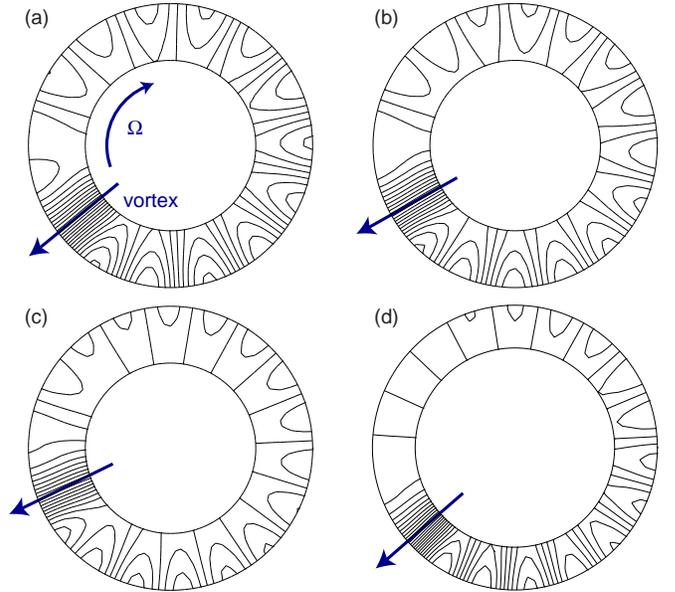, width=1.0\columnwidth}
\caption{Phase profiles $\phi(r,\theta)$ at bias $\gamma$ equal to (a)
$0.5$, (b) $0.4$, (c) $0.3$ on the single vortex resonance of junction
E and at $\gamma = 0.6$ for junction D.  Plotted are lines of constant
phase, their high density corresponds to a large gradient of phase and
hence a large magnetic field.  The position of the vortex is indicated
by an arrow.  The whispering gallery modes with angular wavenumber $k$
equal to (a) $7$, (b) $8$, (c) $9$ and (d) $9$ are observed.}
	\label{fig:phaseSim}
\end{figure}

To confirm the above interpretation of our experimental results, we
performed direct numerical simulations of Eq.~(\ref{eq:2DsineGordon})
in polar coordinates with the boundary conditions
(\ref{eq:2DBoundCond}) using our junction parameters
(see
Table~\ref{tab:parameters}).  We calculated
current-voltage characteristics $V(\gamma) = \Phi_{0} \Omega(\gamma)
\omega_{p} /(2\pi)$ and two-dimensional phase profiles
$\phi(r,\theta,t)$ for various bias points using a plasma frequency of
$\omega_{p}/2\pi = 52.4 \, \rm{GHz}$ determined from
experimental data.  The damping parameter $\alpha = 0.03$ was chosen
close to its estimated experimental value at $T = 4.2 \, \rm{K}$,
$\beta$ here was set to $0$.  The calculated characteristics for
junctions A to E are plotted in Fig.~\ref{fig:IVsim}.  Clearly, the
fine structure on the current-voltage characteristics of the wide
junctions D and E is very well reproduced in the simulation. 
Figure~\ref{fig:phaseSim} shows the phase profiles at bias points on
subsequent fine structure resonances of junction E. Evidently, a clear
whispering gallery structure is found here.  With decreasing bias, the
increase of the angular wave number of the mode from resonance to
resonance is observed.

\begin{figure}[tb]
	\centering
    \epsfig{file=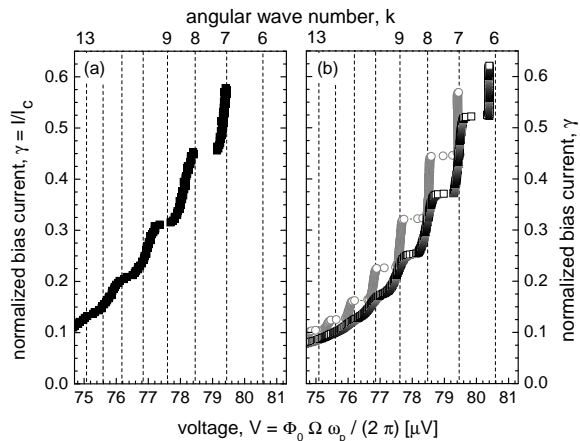, width=3.0 in}
    \caption{(a) Upper part of experimental single-vortex current-voltage
    characteristic of junction E. (b) Simulated current-voltage
    characteristic of this junction for $\alpha = 0.03$, $\beta=0$ (open
    circles) and $\alpha=0.02$, $\beta=0.0012$ (open squares).  The
    calculated resonance voltages are indicated by vertical dotted lines
    and marked by the corresponding wavenumber $k$.}
	\label{fig:expTheoComp17IV}
\end{figure}

Using the resonance condition (\ref{eq:resCond}) and the 
proportionality between the angular frequency of the vortex 
$\Omega$ and the voltage $V$ measured in experiment, the fine structure 
resonances can be fitted according to the formula
\begin{equation}
	V = \Phi_{0} \frac{\omega_{p}}{2\pi} \Omega 
	  = \Phi_{0} \frac{\omega_{p}}{2\pi} \frac{\omega_{k}}{k} \:\: ,
	  \label{eq:freqvolt}
\end{equation}
where $k$ is integer.  In the limit of $\delta \ll 1$ and $k \gg 1$
the linear mode spectrum can be analytically approximated as
$\omega_{k} = r_{e}^{-1} (k + 0.808 \, k^{1/3})$ \cite{Kurin98}. 
More accurately, we have calculated the values of $\omega_{k}$
numerically solving Eq.~(\ref{eq:2DLineareq}) with the boundary
conditions (\ref{eq:2DBoundCond}).
The best fit to the experimental data is found for the plasma
frequency $\omega_{p}/2 \pi = 52.4 \, \rm{GHz}$ and the wavenumber
$k=7$ for the highest voltage resonance (see
Fig.~\ref{fig:expTheoComp17IV}a).  This value of $k$ is exactly the
one found for the highest resonance in numerical simulations (see
Fig.~\ref{fig:phaseSim}a).  The voltages of the resonances found in
simulation (Fig.~\ref{fig:expTheoComp17IV}b) are close to the ones
calculated from the dispersion relation (dotted lines).  The
current-voltage characteristic of junction E simulated for the damping
parameters $\alpha=0.02$ and $\beta=0.0012$ is shown in
Fig.~\ref{fig:expTheoComp17IV}b (open squares).  Clearly, taking into
account the quasiparticle surface losses ($\propto \beta$) leads to a
larger differential resistance of the resonance, which closely
resembles the experimental data in Fig.~\ref{fig:expTheoComp17IV}a.

Considering the resonance condition Eq.~(\ref{eq:resCond}) and the
dispersion of the linear modes $\omega_{k}$, it can be shown that the
density of resonances in voltage and the wave number of the
lowest excited mode increase with decreasing junction width
$w=r_{e}-r_{i}$. This fact was verified by numerical
calculations for junction D, where the lowest mode number excited on
the top of the resonance was found to be $k=9$ (see
Fig.~\ref{fig:phaseSim}d).  For very narrow rings no fine
structure is observed in experiment and in simulation, due to the
overlapping of neighboring resonances in the presence of damping.

The origin of the observed fine structure has also been confirmed to
be due to the interaction of the vortex with the whispering gallery
modes of the junction by investigating its dependence on temperature,
number of trapped vortices, and external magnetic field.  At high
temperatures, no whispering gallery resonances are excited due to the
large intrinsic damping, i.e.~no fine structure is observed. 
Decreasing the temperature below $4.2 \, \rm{K}$, fine structure is
observed in all samples A to E; also the differential resistance of
the resonances decreases with temperature.  Moreover, we found that
the voltages of the fine structure resonances scale with the number
$n$ of moving vortices (or vortex/anti-vortex pairs).  Therefore, for
$n>1$ the fine structure gets clearly resolved in voltage and also
more pronounced, because several vortices coherently pump the
whispering gallery mode.  No dependence of the fine structure step
voltage positions on small external magnetic fields was noticed.  We
have also investigated more narrow annular junctions with a wide idle
region both experimentally and theoretically \cite{wallraff99}.  In
this case, the spectrum of the whispering gallery modes (and, thus, of
the fine structure) is strongly influenced by the geometry and the
electrical properties of the passive region.  The fine structure
recently reported in Ref.~\cite{Martucciello98} appears to be
consistent with our observations.

In summary, we have presented experimental and numerical evidences for
the excitation of whispering gallery modes by vortices moving in wide
annular junctions.  This novel effect has been observed at
sufficiently low damping for annular junctions in a wide range of
electrical and geometrical parameters.  It is very robust with respect
to small external perturbations such as variations in bias current density,
boundary conditions or junction inhomogeneities.  The resonance
frequencies have been calculated and quantitative agreement with
experimental data and numerical simulations better than one percent
has been reached.  Thus, the vortices appear to whisper (generate
radiation) at frequencies between $250$ and $450$ GHz in the annular
whispering gallery of $100 \, \rm{\mu m}$ diameter.

We thank A.~Franz and D.~Kruse for fruitful discussions.  V.~V.~K.~is 
grateful for the support by the Russian Foundation for Basic Research 
(Grant no.~9702-16928) and for partial support by the German Ministry of 
Science and Technology (BMBF Grant no.~13N6945/3).


\end{document}